\documentclass[pra, aps, twocolumn, floatfix, showpacs]{revtex4-1}
\usepackage{graphicx, amsmath, amssymb, times}

\begin{document}
\title{Trapped $^{173}$Yb Fermi gas across an orbital Feshbach resonance}
\author{M. Iskin}
\affiliation{
Department of Physics, Ko\c c University, Rumelifeneri Yolu, 34450 Sar{\i}yer, Istanbul, Turkey.
}
\date{\today}
\begin{abstract}

Starting with the two-band description of an orbital Feshbach resonance, 
we study superfluid properties of a trapped $^{173}$Yb Fermi gas under the 
assumptions of a local-density approximation for the trapping potential and 
a mean-field approximation for the intra-band Cooper pairings. In particular, 
we investigate the competition and interplay between the pair-breaking effect 
that is caused by the inter-band detuning energy, and the pair-breaking and 
thermal-broadening effects that are simultaneously caused by the temperature.
We predict several experimental signatures that are directly caused by this 
interplay including a spatial separation of superfluid and normal phases within 
the trap, and could play decisive roles in probing two-band superfluidity in these 
systems.

\end{abstract}
\maketitle

\section{Introduction}
\label{sec:intro}

Towards the end of last year, two experimental groups have independently 
identified a new type of two-body scattering resonance in an ultracold Fermi 
gas that is composed of neutral $^{173}$Yb atoms~\cite{pagano15, hofer15}. 
The possible creation of the so-called orbital interaction-induced Feshbach 
resonances was proposed a few months earlier as a result of the scattering 
between two (two-electron) alkaline-earth atoms in different electronic-orbital 
and nuclear-spin states~\cite{zhang15, cheng16}. This is in contrast to the more 
familiar magnetic Feshbach resonances, which occur as a result of the 
coupling between two (one-electron) alkali atoms in two different hyperfine 
states~\cite{chin10}. 

It turns out that these distinct resonance mechanisms give rise to important 
implications for the related many-body problems, e.g., in the contexts of
Cooper pairing and associated BCS-BEC evolution~\cite{zhang15, xu16, iskin16}. 
While a single-band description taking only the open-channel scattering 
is typically sufficient for the entire evolution across a magnetic resonance~\cite{giorgini08}, 
a two-band description taking both the open- and closed-channel 
scatterings on an equal footing is minimally required for an orbital 
resonance~\cite{zhang15, cheng16}. Thus, these new systems naturally 
break the ground for studies on two-band superfluidity and intrinsic 
Josephson effect in atomic settings with a high degree of precision and 
control~\cite{iskin16}. In particular, depending on the details of the inter-band 
interactions, one can explore not only the competition between the 
$0$-(in)-phase and $\pi$-(out-of)-phase solutions for the relative phase 
difference between the intra-band superfluid order parameters, but also the 
corresponding relative phase fluctuations and the resultant Gaussian collective 
modes around the equilibrium values, i.e., the phonon-like in-phase Goldstone 
mode and the exciton-like out-of-phase Leggett mode~\cite{iskin16, he16, zhang16}.

Encouraged by the recent realizations of an orbital Feshbach resonance
in a $^{173}$Yb Fermi gas~\cite{pagano15, hofer15}, and unlike the followed 
up theoretical preprints appeared on uniform 
systems~\cite{zhang15, xu16, iskin16, he16, zhang16}, here we focus on 
the confinement-induced signatures that can be decisively traced back 
to the existence of two-band superfluidity in trapped systems. 
For this purpose, we consider a two-band model under the assumptions 
of a local-density approximation for the trapping potential and a 
mean-field approximation for the intra-band Cooper pairings. 
We find that the interplay between the pair-breaking effect that is caused by 
the inter-band detuning energy, and the pair-breaking and thermal-broadening 
effects that are simultaneously caused by the temperature gives rise 
to non-monotonous evolutions in some physical observables. In particular to
the zero temperature, we also find that while the entire trapped gas is a 
superfluid for low detunings, a spatial separation between the central 
superfluid core and the outer normal edge consisting only of particles in the 
lower band eventually appears beyond a detuning threshold that is of the 
order of the resonance value.

The rest of the paper is organized as follows. First, assuming a local-density 
approximation for the trapping potential, we introduce a two-band 
model for the Hamiltonian density in real space in Sec.~\ref{sec:lda}, and 
relate its bare theoretical parameters to the two-body scattering parameters 
of $^{173}$Yb atoms. Then, assuming a mean-field approximation 
for the intra-band Cooper pairings, we derive the mean-field Hamiltonian 
density in Sec.~\ref{sec:mft}, and obtain a set of self-consistency equations 
for the intra-band order parameters and number equations for the two bands. 
Having solved these equations numerically in Sec.~\ref{sec:numerics}
and provided a thorough analysis for our findings, we end the paper with a 
brief summary of our conclusions in Sec.~\ref{sec:conc}. The experimental 
context is briefly discussed in the Appendix.

\section{Local-Density Approximation for Trap}
\label{sec:lda}

The semi-classical method based on a local-density approximation for the
trapping potential is probably one of the most convenient approaches for 
studying many-body effects in finite-sized systems. By shifting the confinement 
potential $V_{tr}(r)$ from the chemical potential $\mu$, one simply introduces 
a local chemical potential $\mu(r) = \mu-V_{tr}(r)$ that depends explicitly 
on the radial distance $r$. This assumption works best for large systems with 
slowly-varying potentials since the relevant Fermi energy scale becomes much 
larger than the confinement-induced energy separation between the quantum 
levels as the number of particles increases. For instance, within this approximation, 
the Hamiltonian density describing isotropically-trapped Fermi gases across 
an orbital Feshbach resonance can be written as~\cite{iskin16}
\begin{align*}
H(r) = \sum_{i \sigma \mathbf{k}} \xi_{i \mathbf{k}}(r) c_{i \sigma \mathbf{k}}^\dagger(r)  c_{i \sigma \mathbf{k}}(r)
- \sum_{i j \mathbf{q}} V_{ij} b_{i \mathbf{q}}^\dagger(r) b_{j \mathbf{q}}(r),
\label{eqn:ham}
\end{align*}
where the band index $i \equiv \{1, 2\}$ refers to the particles in the 
open (lower band) and closed (upper band) channels with pseudo-spin 
projections $\sigma \equiv \{\uparrow, \downarrow\}$, and $\mathbf{k}$ is momentum. 
The operator $c_{i \sigma \mathbf{k}}^\dagger(r)$ creates a single particle at $r$ 
with quantum numbers $i$, $\sigma$ and $\mathbf{k}$, and dispersion
$
\xi_{i \mathbf{k}}(r) = \varepsilon_{\mathbf{k}} - \mu_i(r).
$
Here, $\varepsilon_{\mathbf{k}} = k^2/(2m)$ is in units of $\hbar = 1$, and
$\mu_1(r) = \mu - V_{tr}(r)$ is for the lower and $\mu_2(r) = \mu - \delta/2 - V_{tr}(r)$
is for the upper band, where $V_{tr}(r) = m \omega^2 r^2/2$ is assumed to be 
harmonic in space, and the energy shift $\delta/2 \ge 0$ between the two bands 
is a controllable detuning parameter that is used to access an orbital Feshbach 
resonance. Similarly, the operator
$
b_{i \mathbf{q}}^\dagger(r) = \sum_{\mathbf{k}} 
c_{i \uparrow, \mathbf{k}+\mathbf{q}/2}^\dagger(r)  
c_{i \downarrow, -\mathbf{k}+\mathbf{q}/2}^\dagger(r)
$
creates pairs of $\uparrow$ and $\downarrow$ particles at $r$ with quantum numbers 
$i$ and center-of-mass momentum $\mathbf{q}$. 
The bare amplitudes for the local intra-band
$
V_{11} = V_{22} = (g_- + g_+)/2
$
and local inter-band
$
V_{12} = V_{21} = (g_- - g_+)/2
$
interactions are related to the two-body scattering lengths in vacuum $a_{s\pm}$ 
via the usual renormalization relations
$
1/g_\pm = -m \mathcal{V}/(4\pi a_{s\pm}) + \sum_{\mathbf{k}} m/k^2,
$
where $\mathcal{V}$ is the volume, in such a way that the orbital resonance occurs 
precisely when $\delta$ is tuned to a critical threshold 
$\delta_{res} = 4/[m (a_{s-} + a_{s+})^2]$~\cite{zhang15}.
These parameters have recently been determined for a $^{173}$Yb Fermi gas, 
and are given by $a_{s+} \approx 1900a_0$ and $a_{s-} \approx 200a_0$
with $a_0$ the Bohr radius~\cite{pagano15, hofer15}, for which both 
intra- and inter-band interactions turned out to be attractive with $V_{ij} > 0$.

\section{Mean-Field Approximation for Pairing}
\label{sec:mft}

Assuming that the fluctuations of the pair-creation operators are small in 
comparison to their equilibrium values, we adopt a mean-field approximation 
for pairing, and introduce an intra-band order parameter
$
\Delta_{i\mathbf{q}}(r) = - \sum_j V_{ij} \langle b_{j \mathbf{q}}(r) \rangle
$
for each band~\cite{iskin16}, where $\langle \cdots \rangle$ is a thermal average. 
In addition, restricting ourselves solely to local BCS-like solutions, we set 
$\mathbf{q} = \mathbf{0}$ and determine the local complex parameter 
$\Delta_i(r) = \Delta_{i\mathbf{0}}(r)$ self-consistently with the corresponding 
local number equation
$
n_i(r) = \sum_{\sigma \mathbf{k}}
\langle c_{i \sigma \mathbf{k}}^\dagger(r)  c_{i \sigma \mathbf{k}}(r) \rangle
$
for each band at a given $r$. Once the total number of particles in a given band 
is obtained by
$
N_i = (1/\mathcal{V}) \int d^3\mathbf{r} n_i(r)
$
then $\mu$ is iterated until $N = N_1 + N_2$ is fixed to a specified value
given in Sec.~\ref{sec:numerics}. This self-consistent construction is a 
straightforward extension of the usual mean-field approach that has 
extensively been employed for single-band Fermi gases, and it forms 
the fundamental basis for most of the BCS-BEC crossover studies in the 
literature, over the past decade or so, in the context of magnetic Feshbach 
resonances~\cite{giorgini08}.

Thus, within such a mean-field approximation for the intra-band pairings, 
the local mean-field Hamiltonian can be reexpressed as
\begin{align}
H_{mf}(r) &= \sum_{i \sigma \mathbf{k}} \xi_{i \mathbf{k}}(r) c_{i \sigma \mathbf{k}}^\dagger(r)  c_{i \sigma \mathbf{k}}(r) \nonumber \\ 
& + \sum_{i\mathbf{q}} \left[ \Delta_{i\mathbf{q}}(r) b_{i \mathbf{q}}^\dagger(r) 
+  \Delta_{i\mathbf{q}}^*(r) b_{i \mathbf{q}}(r) \right] \nonumber \\
& + \sum_{ij\mathbf{q}} U_{ij} \Delta_{i\mathbf{q}}^*(r) \Delta_{j\mathbf{q}}(r),
\end{align}
where the matrix $\mathbf{U}$ is the inverse of the amplitude matrix $\mathbf{V}$,
i.e., its elements can be written explicitly as
$U_{11} = V_{22}/\det\mathbf{V}$,
$U_{22} = V_{11}/\det\mathbf{V}$,
$U_{12} = -V_{12}/\det\mathbf{V}$ and
$U_{21} = -V_{21}/\det\mathbf{V}$
with $\det\mathbf{V} = V_{11}V_{22}-V_{12}V_{21}$. Note that the inter-band 
coupling gives rise to a Josephson-type contribution to the Hamiltonian, 
$
U_{12} (\Delta_{1\mathbf{q}}^* \Delta_{2\mathbf{q}} 
+ \Delta_{1\mathbf{q}} \Delta_{2\mathbf{q}}^*),
$
depending explicitly on the relative phase between the intra-band order parameters.
Then, restricting to local BCS-like solutions, the resultant self-consistency 
equations can be compactly put in a more familiar form as follows~\cite{iskin16}
\begin{align}
\label{eqn:op}
\Delta_i(r) & = \sum_{j \mathbf{k}} V_{ij} \frac{\Delta_j(r)}{2E_{j \mathbf{k}}(r)} \tanh \left[\frac{E_{j \mathbf{k}}(r)}{2T} \right], \\
\label{eqn:ne}
n_i(r) & = \sum_{\mathbf{k}} \left\lbrace 1 - \frac{\xi_{i \mathbf{k}}(r)}{E_{i \mathbf{k}}(r)} \tanh \left[\frac{E_{i \mathbf{k}}(r)}{2T} \right] \right\rbrace,
\end{align}
where
$
E_{i \mathbf{k}}(r) = \sqrt{\xi_{i \mathbf{k}}^2(r) + |\Delta_i(r)|^2}
$
is the energy of the local quasi-particle excitations in the $i$th band with 
momentum $\mathbf{k}$, $T$ is the temperature, and the Boltzmann 
constant $k_B$ is set to unity. The summand in Eq.~(\ref{eqn:ne}) is the
local momentum distribution $n_i(r, \mathbf{k})$ of particles in the $i$th band.

Motivated by the success of the analogous mean-field theories in describing 
the fundamental properties of alkali atoms across a magnetic Feshbach 
resonance~\cite{giorgini08}, here we apply it to alkaline-earth atoms across 
an orbital Feshbach resonance. Therefore, we are interested in the so-called 
$\pi$-phase solution for the local relative phases between the local order 
parameters, i.e., $\textrm{sign} [\Delta_1(r)] = - \textrm{sign} [\Delta_2(r)]$ 
at any given $r$, which is directly linked to the orbital Feshbach resonance 
found in a $^{173}$Yb Fermi gas~\cite{zhang15, iskin16}.

\section{$\pi$-Phase Solutions for a $^{173}$Yb Fermi gas}
\label{sec:numerics}

We use the following definitions of an effective Fermi energy and the associated 
Fermi momentum $\varepsilon_F = k_F^2/(2m)$, and the corresponding 
Thomas-Fermi radius $r_F$ in presenting our numerical solutions. 
Assuming a total of $N$ non-interacting particles in a single-band Fermi 
gas at $T = 0$, and setting $\mu = \varepsilon_F$, we may write
$
\varepsilon_F = k_F^2(r)/(2m) + m \omega^2 r^2/2
$ 
for the lower band within the local-density approximation. 
This defines a local Fermi momentum $k_F(r)$ in such a way that the local 
number of particles is given by $n(r) = V k_F^3(r)/(3\pi^2)$ at a given $r$. 
Noting that $k_F(r_F) = 0$ at the edge of the system by definition, we may express 
$k_F = k_F(0) = m \omega r_F$, leading to $N = k_F^3 r_F^3 /24$ or 
equivalently $\varepsilon_F = \omega (3N)^{1/3}$. Choosing a typical atomic
density $n(0)/\mathcal{V} \approx 10^{14} \textrm{cm}^{-3}$ at the center 
of the trap and using the scattering parameters of a $^{173}$Yb Fermi gas 
given in Sec.~\ref{sec:lda}, we find $1/(k_F a_{s+}) \approx 0.693$, 
$1/(k_F a_{s-}) \approx 6.582$ and $\delta_{res} \approx 3.144\varepsilon_F$.
In addition, by choosing a large momentum cut-off $k_0 = 100k_F$ in 
$\mathbf{k}$-space sums, we obtain $k_0$-independent solutions for the 
physical observables, even though all of the bare interaction amplitudes 
$V_{ij}$ themselves depend explicitly on $k_0$.

\begin{figure}[htb]
\centerline{\scalebox{0.6}{\includegraphics{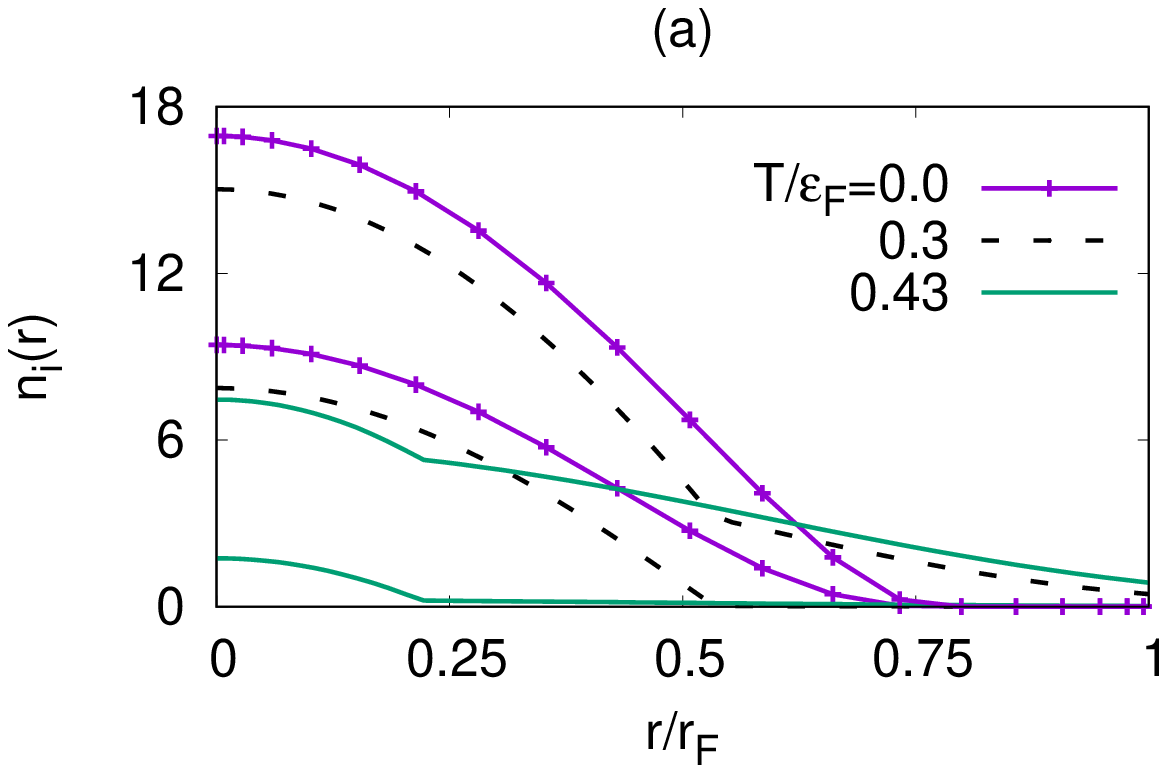}}}
\centerline{\scalebox{0.6}{\includegraphics{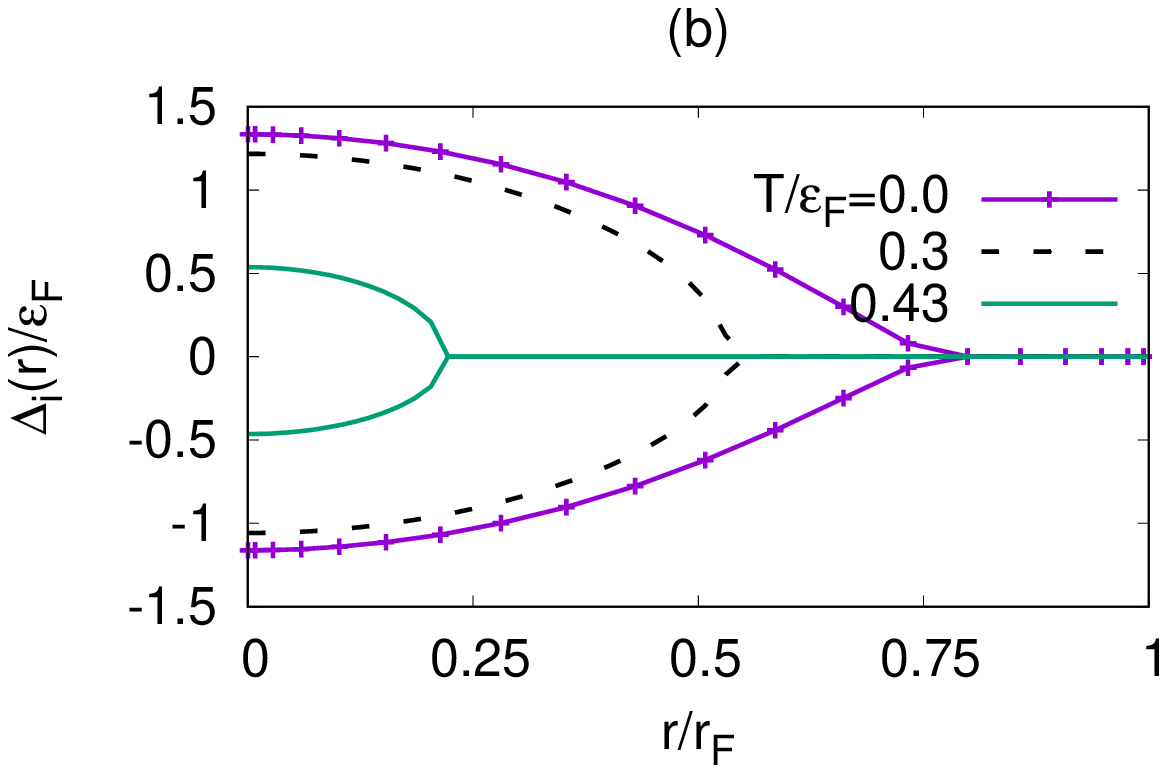}}}
\centerline{\scalebox{0.6}{\includegraphics{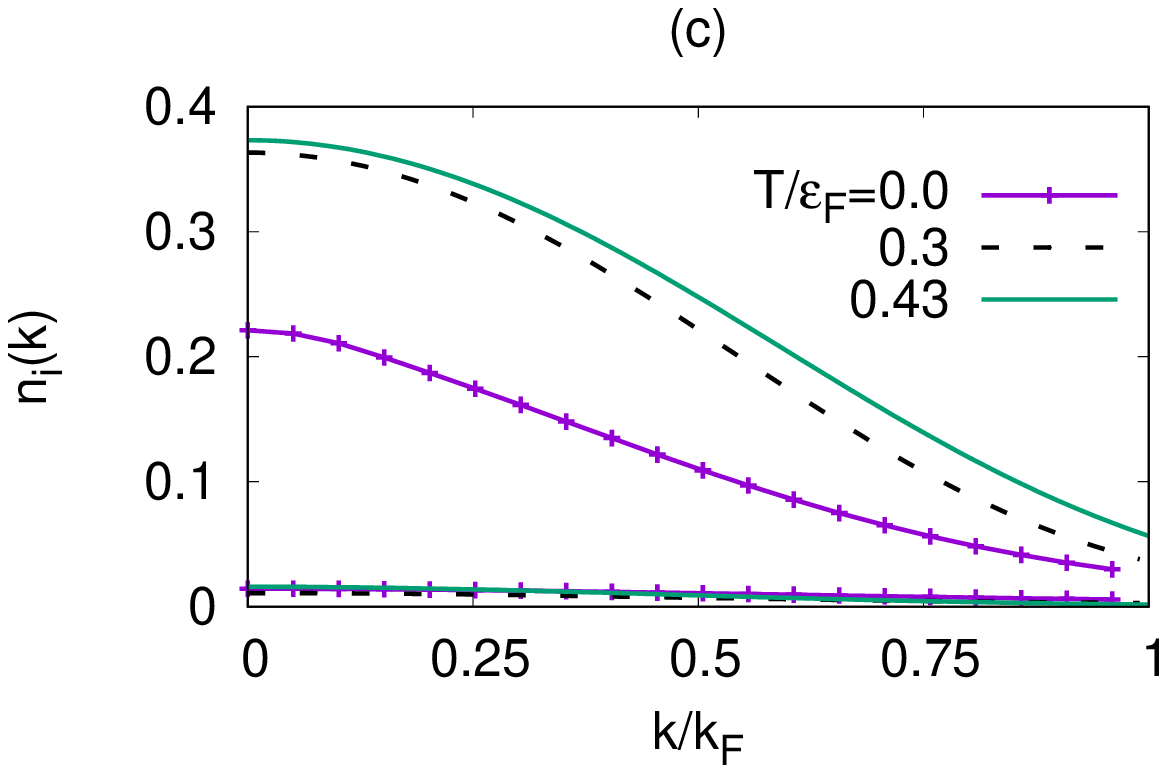}}}
\caption{\label{fig:rad} (Color online) 
\textit{Radial profiles at resonance detuning}.
(a) The numbers of particles $n_i(r)$ [in units of $N\mathcal{V}/(4\pi r_F^3)$], and
(b) the order parameters $\Delta_i(r)$ 
are shown as functions of the radial distance $r$.
(c) The trap-averaged momentum distributions $n_i(k)$ are shown in units of 
$4\pi r_F^3/\mathcal{V}$ and as functions of the radial momentum $k$.
Here, $i = \{1, 2\}$ corresponds, respectively, to the lower and upper band, 
where $n_1(r) > n_2(r)$ in (a), $|\Delta_2(r)| > |\Delta_1(r)|$ and $\Delta_1(r) < 0$ 
in (b), and $n_1(k) > n_2(k)$ in (c).
}
\end{figure}

First we consider a resonant Fermi gas with $\delta = \delta_{res}$, and 
present typical $n_i(r)$ and $\Delta_i(r)$ profiles as functions of $r$. 
It is worth mentioning here that since were are presenting the $\pi$-(out-of)-phase 
excited-state solutions but not the $0$-(in)-phase ground-state ones, 
the higher $i = 2$ band has higher order parameters in spite of its lower 
density. As shown in Figs.~\ref{fig:rad}(a) and~\ref{fig:rad}(b), while 
$|\Delta_2(r)| > |\Delta_1(r)| > 0$ as long as $n_1(r) > n_2(r) > 0$ at $T = 0$, 
and therefore, the entire gas is found to be a superfluid, the pair-breaking 
effect caused by finite $T$ weakens $|\Delta_i(r)|$ and turns the edge of the 
gas to normal beyond a critical radius $r > r_S$. Here, the critical radius 
$r_S$ for the spatial separation of superfluid and normal phases within the trap is 
determined by the simultaneous vanishing of $|\Delta_{1,2}(r_S^-)| \to 0^+$. 
Increasing $T$ gradually decreases $r_S$ towards the center of the trap, 
and eventually the entire gas turns to normal, i.e., $r_S \to 0$, 
beyond the critical superfluid-normal transition temperature 
$T_c \approx 0.45 \varepsilon_F$. The simultaneous disappearance of the
order parameters leads not only to observable cusps in $n_i(r)$ precisely 
at $r = r_S$ but also to the thermal broadening of the outer normal regions. 
This is best seen in Fig.~\ref{fig:rad}(c), where we present the trap-averaged 
momentum distributions
$
n_i(k) = (1/\mathcal{V}) \int d^3 \mathbf{r} n_i(r, \mathbf{k})
$
as functions of $k$, where $n_i(r, \mathbf{k})$ is the summand of 
Eq.~(\ref{eqn:ne}).

We note the following in passing for the radial profiles at $T = 0$. 
Up until $\delta \sim \delta_{res}$, the local occupation of the upper 
band in the trap turns out to be non-zero as long as the lower band 
is also locally occupied there, i.e. if $n_1(r) \ne 0$ then $n_2(r) \ne 0$ for 
any given $r$. This is a direct result of the inter-band coupling, and the entire 
gas is a superfluid with $\Delta_2(r) \ne 0$ wherever $\Delta_1(r) \ne 0$, 
as illustrated above for a resonant Fermi gas. On the other hand, 
when $\delta \gtrsim 4\varepsilon_F$, we find that the inter-band coupling is 
not locally strong enough to overcome the detuning barrier towards 
the edge of the gas, as a consequence of which the intra-band pairings 
vanish $|\Delta_{1,2}(r \to r_S)| \to 0^+$ simultaneously at some critical 
radius $r_S$. This naturally gives rise to $n_2(r) = 0$ and $n_1(r) \ne 0$ 
for $r > r_S$, and hence, a spatial separation appears between the central 
superfluid core and the outer normal edge consisting only of particles 
in the lower band. When $r_S$ eventually reduces to $0$ as 
$\delta \gg \varepsilon_F$ then the entire trap is effectively occupied by 
a single-band of non-interacting Fermi gas in the lower band.

\begin{figure}[htb]
\centerline{\scalebox{0.7}{\includegraphics{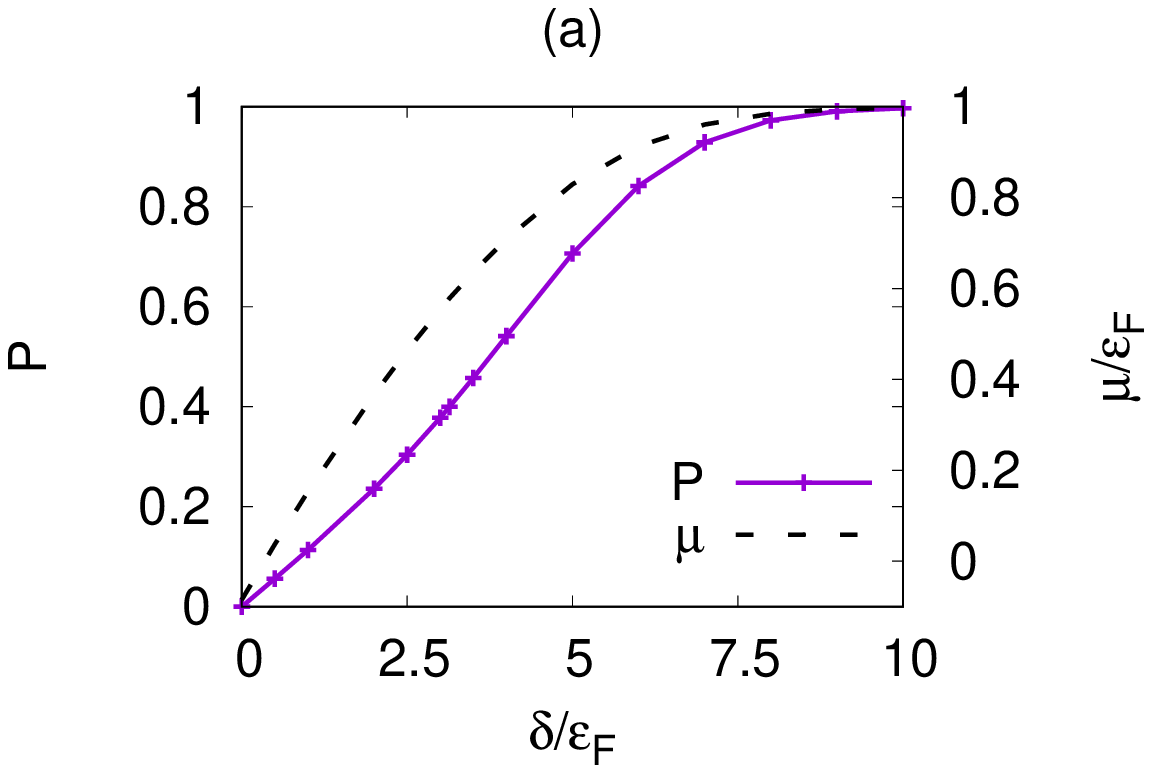}}}
\centerline{\scalebox{0.7}{\includegraphics{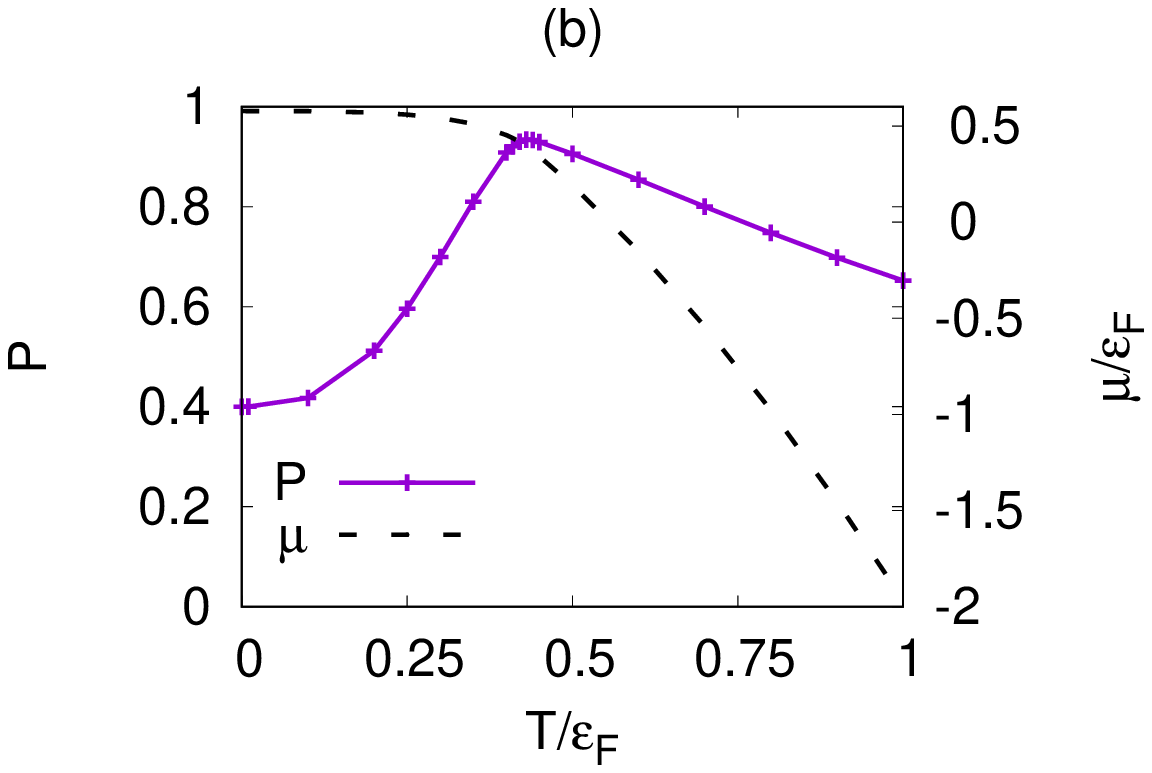}}}
\caption{\label{fig:Pmu} (Color online) 
The band-population imbalance $P = (N_1-N_2)/N$ and chemical 
potential $\mu$ are shown as functions of 
(a) detuning $\delta$ at zero temperature, and
(b) temperature $T$ at resonance detuning.
}
\end{figure}

To understand the general trends, next we present the band-population 
imbalance $P = (N_1-N_2)/N$ and $\mu$ in Fig.~\ref{fig:Pmu}(a) as 
functions of $\delta$ at $T = 0$ . It is clearly seen that while $P = 0$ or 
$N_1 = N_2$ and $\mu < 0$ at $\delta = 0$, the particles gradually transfer 
from the upper to the lower band as a result of the increased energy 
difference $\delta/2$ between the bands and its pair-breaking effect, 
leading eventually to $P \to 1$ or $N_1 \gg N_2 \to 0$ and 
$\mu \to \varepsilon_F$ in the $\delta \gg \varepsilon_F$ limit. 
The evolutions of $P$ and $\mu$ are smooth and monotonous across 
the resonance, at which point we find $P \approx 0.400$ and 
$\mu \approx 0.578\varepsilon_F$. Similarly, in Fig.~\ref{fig:Pmu}(b), 
we present $P$ and $\mu$ as functions of $T$ at $\delta = \delta_{res}$. 
While $\mu$ is a monotonically decreasing function of $T$, $P$ first increases 
to a peak value of $0.934$ at $T \approx 0.43 \varepsilon_F$ and then decreases. 
This temperature almost coincides with the critical one where 
$\mu \approx 0.342 \varepsilon_F$ and $P \approx 0.930$ at 
$T_c \approx 0.45 \varepsilon_F$. 

\begin{figure}[htb]
\centerline{\scalebox{0.6}{\includegraphics{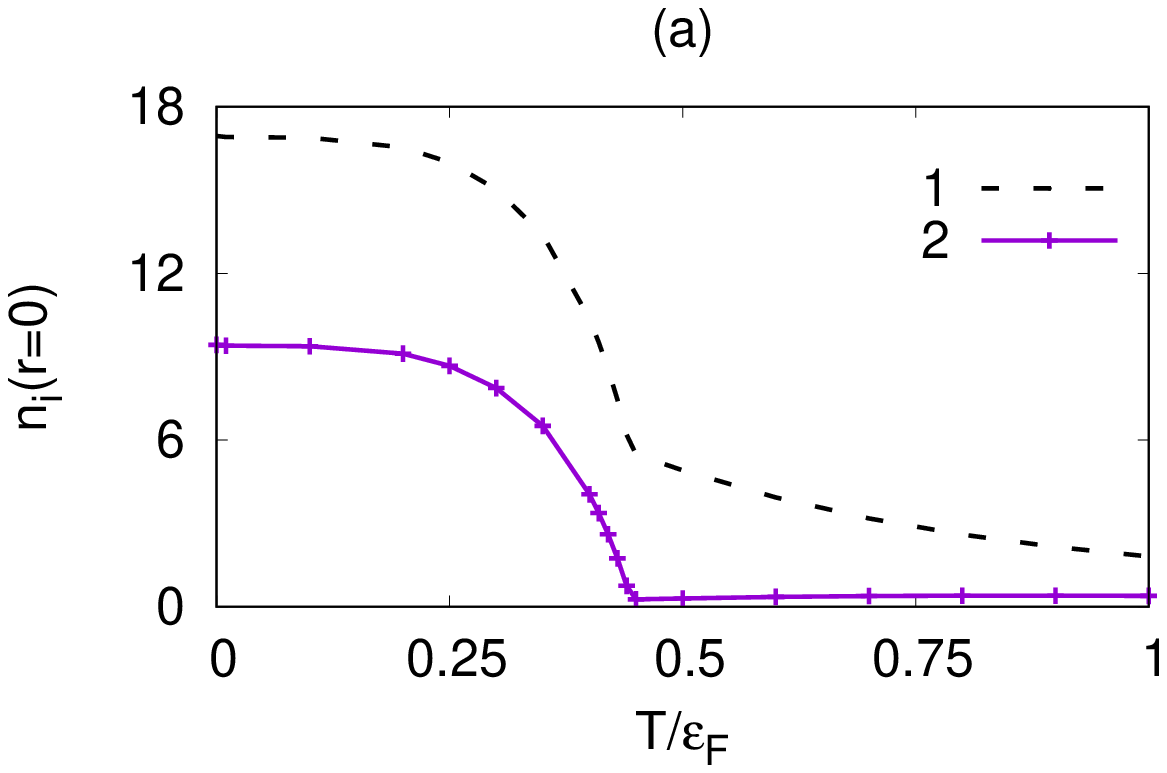}}}
\centerline{\scalebox{0.6}{\includegraphics{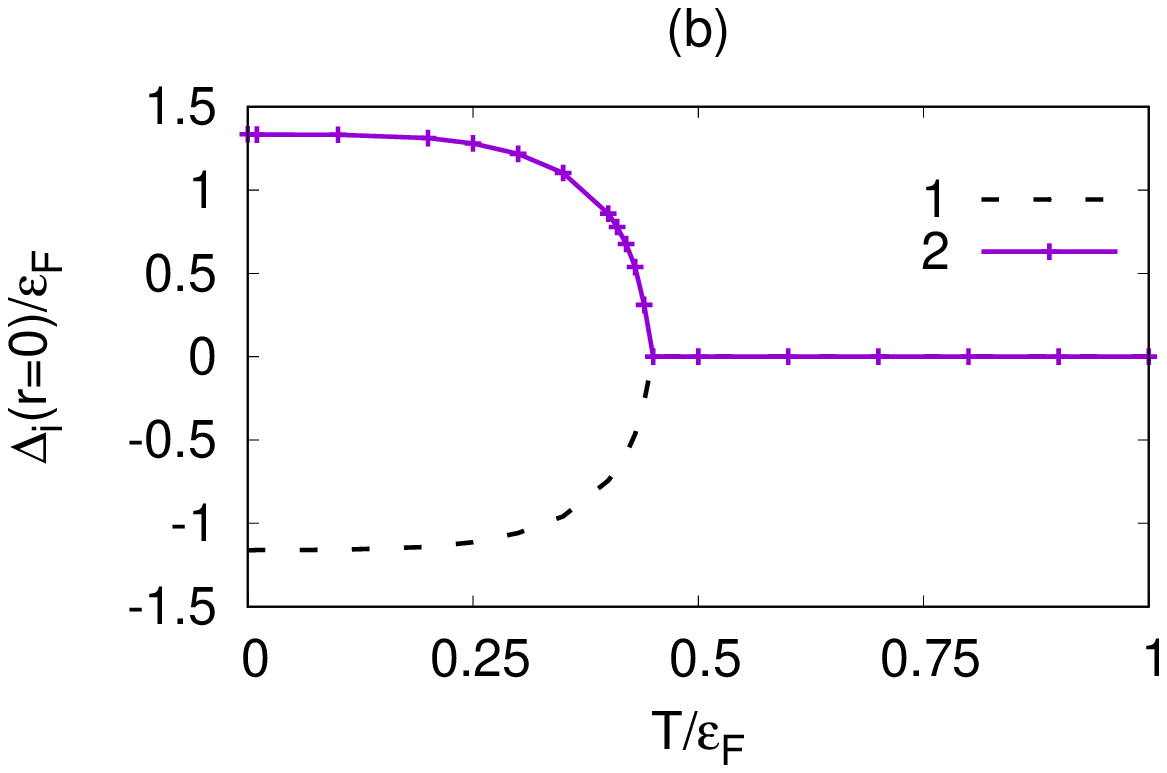}}}
\centerline{\scalebox{0.6}{\includegraphics{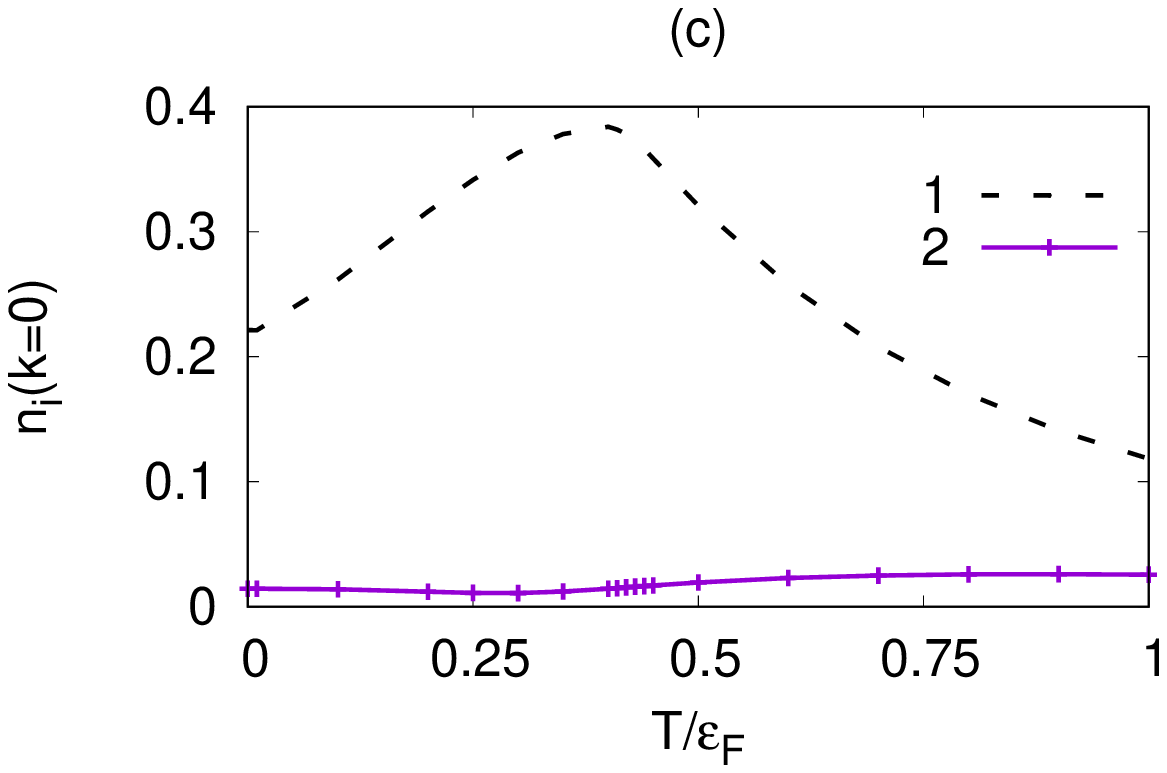}}}
\caption{\label{fig:res} (Color online) 
\textit{Central parameters at resonance detuning}.
(a) The numbers of particles $n_i(r = 0)$ [in units of $N\mathcal{V}/(4\pi r_F^3)$],
(b) the order parameters $\Delta_i(r = 0)$, and
(c) the trap-averaged momentum distributions $n_i(k = 0)$ (in units of $4\pi r_F^3/\mathcal{V})$
are shown as functions of temperature $T$.
}
\end{figure}

The non-monotonous evolution of $P$ with $T$ at fixed $\delta$ is a direct 
consequence of the competition between the pair-breaking and 
thermal-broadening effects of $T$. 
To illustrate this competition, we present the central parameters $n_i(r = 0)$ 
and $\Delta_i(r = 0)$ in Figs.~\ref{fig:res}(a) and~\ref{fig:res}(b), respectively, 
as functions of $T$ for a resonant Fermi gas. In accordance with our definition 
of $\varepsilon_F$ given above for a non-interacting single-band Fermi gas 
at $T = 0$, the upper band is completely empty for $\varepsilon_F < \delta/2$. 
Since $\delta_{res} \approx 3.144\varepsilon_F$ in this paper, $\Delta_i(r) \ne 0$ 
promotes some of the particles to the upper band causing $N_2 \ne 0$ 
at $T = 0$ in the first place, and thus, the reduction of $|\Delta_i(r)|$ at finite 
but low $T \lesssim T_c$ naturally demotes particles back to the lower band. 
However, in the mean time, the particles are thermally excited back to the 
upper band as well, leading to the aforementioned competition as a 
function of $T$. The isolated effects of pair-breaking and thermal-broadening 
mechanisms on the occupations of the bands are evidently seen in 
Fig.~\ref{fig:res}(c), where we present $n_i(k = 0)$ as functions of $T$.

\begin{figure}[htb]
\centerline{\scalebox{0.6}{\includegraphics{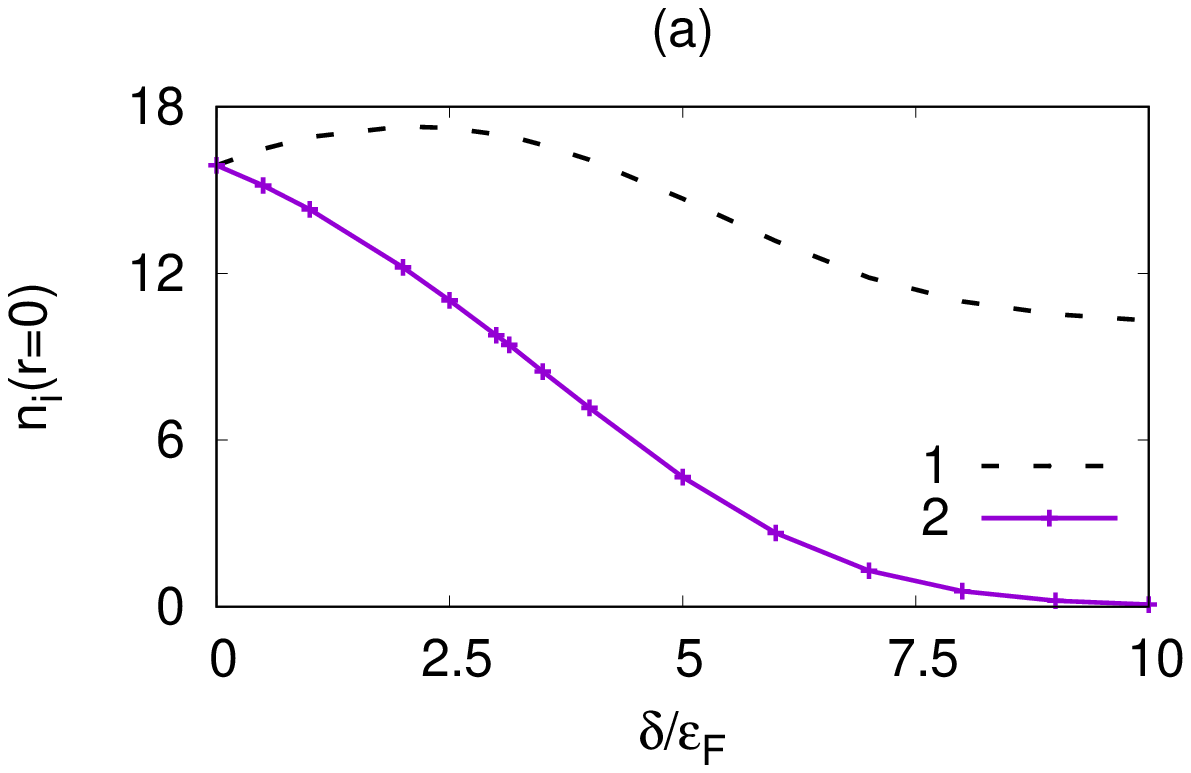}}}
\centerline{\scalebox{0.6}{\includegraphics{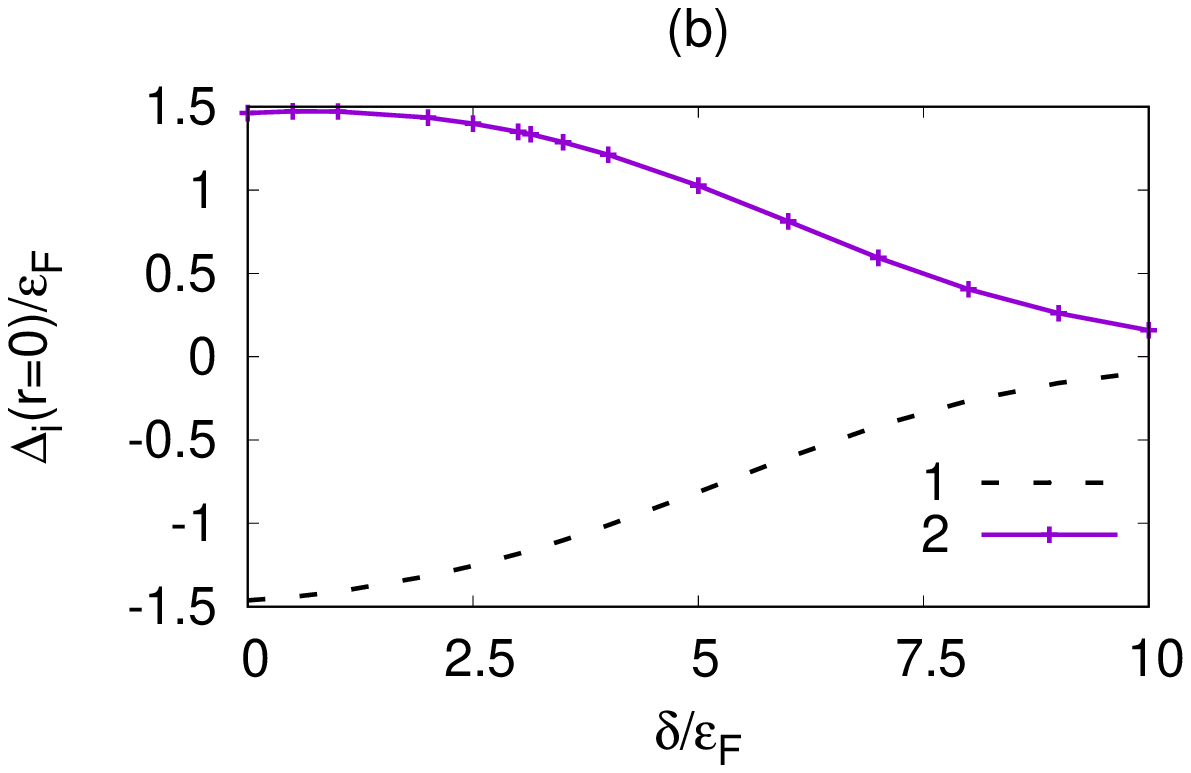}}}
\centerline{\scalebox{0.6}{\includegraphics{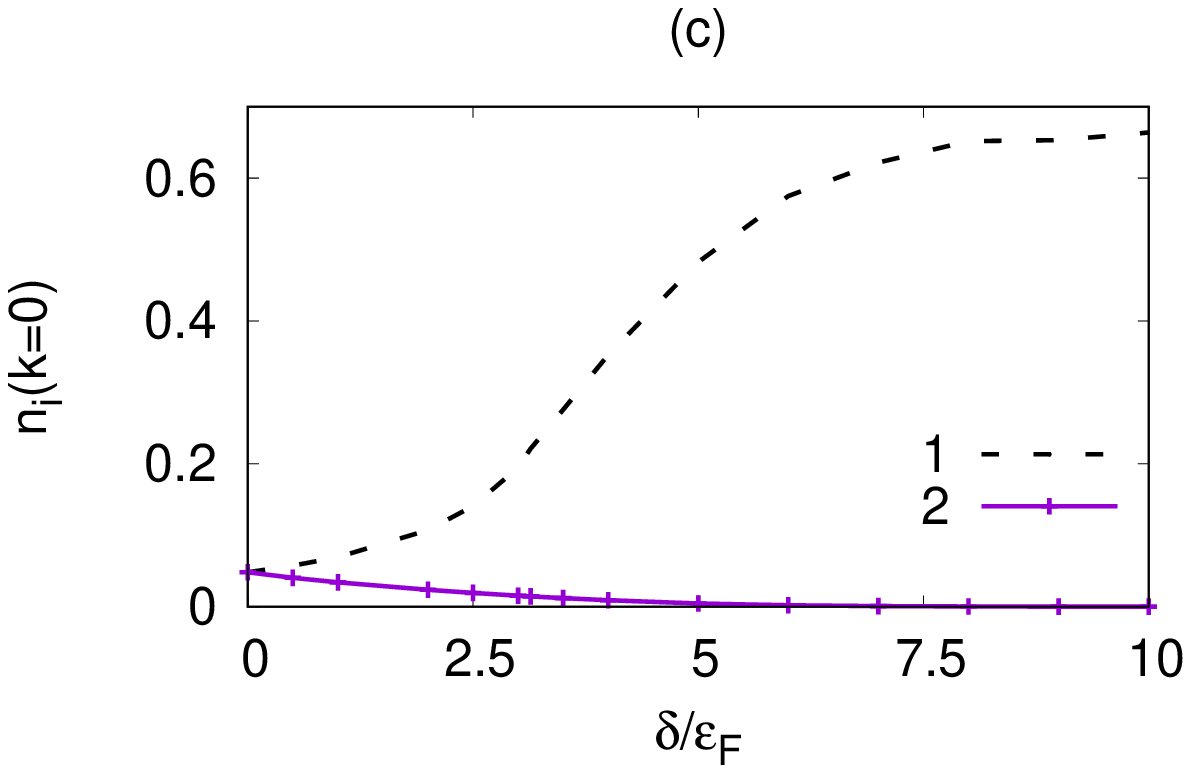}}}
\caption{\label{fig:T0} (Color online) 
\textit{Central parameters at zero temperature}.
(a) The numbers of particles $n_i(r = 0)$ [in units of $N\mathcal{V}/(4\pi r_F^3)$],
(b) the order parameters $\Delta_i(r = 0)$, and
(c) the trap-averaged momentum distributions $n_i(k = 0)$ (in units of $4\pi r_F^3/\mathcal{V})$
are shown as functions of detuning $\delta$.
}
\end{figure}

For completeness, next we discuss the central parameters $n_i(r = 0)$, 
$\Delta_i(r = 0)$ and $n_i( k =0)$ as functions of $\delta$ at $T = 0$, 
showing purely the crucial role played by the pair-breaking effect of 
$\delta$ in the absence of thermal effects. As shown in Fig.~\ref{fig:T0}, 
while $n_1(0) = n_2(0)$ and $|\Delta_1(0)| = |\Delta_2(0)|$ at $\delta = 0$, 
the particles gradually transfer from the upper to the lower band with 
increased $\delta$ due to the simultaneous reduction of 
$|\Delta_2(0)| > |\Delta_1(0)|$. This eventually leads to $n_1(0) \gg n_2(0) \to 0$ 
and $|\Delta_{1,2}(0)| \to 0$ in the $\delta \gg \varepsilon_F$ limit, 
and the problem reduces effectively to a single-band of non-interacting 
Fermi gas in the lower band. 

We would like to remark here that the physical picture outlined just 
above in understanding the general trends presented in this paper 
goes beyond the simple mean-field approximation that is assumed in 
our numerical calculations. It is widely believed that while this approximation 
reliably describes the low-temperature ($T \ll T_c$) properties 
of a weakly-interacting Fermi gas in general, the inclusion of (at least) 
the Gaussian pair-fluctuations is necessary in order to produce a 
qualitatively correct $T_c$ in the strongly-interacting regime especially 
near the resonance~\cite{xu16}. However, the non-monotonous 
evolutions caused by the competition between the pair-breaking and 
thermal-broadening mechanisms should be manifested in beyond 
mean-field calculations as well, apart from expected minor 
quantitative differences.

\section{Conclusions}
\label{sec:conc}

In summary, we analyzed how a trapped $^{173}$Yb Fermi gas and its 
superfluid properties evolve across an orbital Feshbach resonance. We 
used a two-band description for this purpose, under the assumptions of a 
local-density approximation for the trapping potential and a mean-field 
approximation for the intra-band pairings. One of our primary findings 
is that the interplay between the pair-breaking effect that is caused by 
the inter-band detuning energy $\delta$, and the pair-breaking and 
thermal-broadening effects that are simultaneously caused by the 
temperature $T$ gives rise to non-monotonous evolutions in some 
physical observables, including the band-population imbalance 
and trap-averaged momentum distributions. In addition, we found at $T = 0$ 
that while the entire trapped gas is a superfluid for $\delta \lesssim \delta_{res}$ 
with the resonance detuning $\delta_{res} \sim 3\varepsilon_F$, a spatial 
separation between the central superfluid core and the outer normal edge 
which consists only of particles in the lower band eventually appears beyond 
a critical detuning that is of the order of $\delta \gtrsim 4\varepsilon_F$. 
We also argued that, since these predictions are physically intuitive and not 
caused by any of the approximations used, they may play decisive roles in 
probing two-band superfluidity in the cold-atom context. As an immediate 
outlook, we look forward to further research along these lines by especially 
taking the beyond local-density and/or mean-field corrections into account 
for quantitatively more accurate predictions.

\section{Acknowledgments} 
\label{sec:ack}

This work is supported by the funding from T\"{U}B$\dot{\mathrm{I}}$TAK Grant 
No. 1001-114F232 and the BAGEP award of the Turkish Science Academy.

\appendix

\section{Experimental Context}
\label{appsec:exp}

First of all, we consider two different nuclear-spin states of a $^{173}$Yb atom and denote them 
with $|\Uparrow\rangle$ and $|\Downarrow\rangle$. In addition, assuming that a $\pi$-polarized
clock laser light can be used to excite the atoms from their ground ($^1$S$_0$) state to 
a long-lived metastable ($^3$P$_0$) one, we also take into account two different 
internal-orbital states and denote them with $|g\rangle$ and $|e\rangle$~\cite{pagano15, hofer15}.
Then, in Eq.~(\ref{eqn:ham}), the pseudo-spin projections $\sigma$ correspond precisely to
$
|1 \uparrow \rangle \equiv |e \Uparrow \rangle 
$
and
$
|1 \downarrow \rangle = |g \Downarrow \rangle
$
in the lower ($i = 1$) band, and to
$
|2 \uparrow \rangle = |g \Uparrow \rangle
$
and
$
|2 \downarrow \rangle = |e \Downarrow \rangle
$
in the upper ($i = 2$) band. This reorganization is in such a way that the nuclear-spin 
projections and orbital states are directly linked with each other in the two-particle 
scattering channels, where the anti-symmetric state
$
|e \Uparrow; g \Downarrow \rangle = 
(| e \Uparrow \rangle | g \Downarrow \rangle - | g \Downarrow \rangle | e \Uparrow \rangle)/\sqrt{2}
$
corresponds to the open channel, and
$
|g \Uparrow; e \Downarrow \rangle 
= (| g \Uparrow \rangle | e \Downarrow \rangle - | e \Downarrow \rangle | g \Uparrow \rangle)/\sqrt{2}
$
to the closed one~\cite{zhang15,cheng16}.

Since the two-particle interaction between one $|g\rangle$ atom and one $|e\rangle$
atom in two different nuclear-spin states is characterized by the interplay between 
the orbital-singlet scattering length ($a_{s+}$) and the orbital-triplet ($a_{s-}$) one given
in the main text, it is possible to have both intra-channel spin-conserving (direct) interactions 
as well as an inter-channel spin-flipping (exchange) one. That is, the interaction 
between one $|g\rangle$ and one $|e\rangle$ atom may also involve a spin-flip. 
While the strengths of the former are equally proportional to an effective 
direct scattering length $({a_{s+}} + a_{s-})/2$ in both open and closed channels, 
that of the latter one is proportional to an effective exchange scattering length 
$({a_{s+}} - a_{s-})/2$ giving rise to a coupling between the open and closed channels 
when $a_{s+} \ne a_{s-}$. 

Furthermore, the presence of an external magnetic field splits the nuclear-spin states 
depending on their Zeeman level, shifting relatively the energies of the scattering 
channels by varying the strength of the field. For instance, a strong 
magnetic field weakens the coupling between open and closed channels as the 
Zeeman energy dominates the spin-exchange interactions leading to well-defined 
nuclear-spin states. Therefore, the two-particle scattering channels may be strongly 
correlated with each other at small and intermediate magnetic fields, allowing for the 
creation of a new type of magnetically-tunable orbital Feshbach resonance, 
once the Zeeman energy matches the two-body binding energy of the least bound 
state in the closed channel~\cite{zhang15,cheng16,pagano15, hofer15}. 

As the effective nuclear magnetic moments involved in orbital resonances are much 
smaller than the electronic ones in alkali-atom resonances, the widths of these
resonances can be broad in magnetic field, despite their large and negative effective 
ranges which are characteristic features of narrow, i.e., closed-channel dominated, 
alkali-atom resonances. Therefore, in contrast to the broad alkali-atom resonances 
where it is sufficient to retain only the open channel with a single order parameter
as the minimal description of the BCS-BEC crossover physics, here it is necessary 
to treat open and closed channels on an equal footing by introducing a coupled set
of two mutually-coherent order parameters requiring a self-consistent solution, 
as discussed in the main text.

Lastly, we restrict ourselves to the balanced number of $\uparrow$ and $\downarrow$ 
atoms in each band for its simplicity, and set $\mu_\uparrow = \mu_\downarrow = \mu$ in
Eq.~(\ref{eqn:ham}). This is such that the total number $N$ of atoms are equally 
distributed over the two states $|e \Uparrow \rangle$ and $|g \Downarrow \rangle$ of the 
open channel in the non-interacting limit when $\delta > 2\epsilon_F$. For instance, 
if all of the atoms are initially prepared in the ground state $|g \rangle$ then one can 
achieve a balanced system by exciting all of the $\Uparrow$ atoms from $|g\rangle$
to the excited state $|e\rangle$. The formalism developed in this paper can easily 
be extended to the analysis of the imbalanced problem, and this is one of the 
immediate experimental interests to be addressed in the near future. Furthermore,
we assume a common trapping potential for all atoms, independent of their orbital 
and nuclear-spin degrees of freedom~\cite{pagano15, hofer15}.

\end{document}